\documentclass[12pt,a4paper,dvips]{article}
\usepackage{a4p}
\usepackage{cite,mcite}
\usepackage{graphicx}
\usepackage{epsfig}
\usepackage{l3_title,ifthen}
\usepackage{physics,Lep}
\usepackage{a4p}
\date{September 20, 1999}
\preprint{99-128} 
\def\dm        {\ensuremath{\Delta M}}
\def\susy#1{\ensuremath{\tilde{\mathrm{#1}}}}%
\def\slepton   #1{{\susy{\ell}^{#1}}}
\def\selectron #1{\ensuremath{\susy{e}^{#1}}}
\def\smuon     #1{\ensuremath{\susy{\mu}^{#1}}}
\def\stau      #1{\ensuremath{\susy{\tau}^{#1}}}

\def\neutralino#1{\ensuremath{\susy{\chi}_{#1}^0}}
\def\EV30{\ensuremath{E_{\mathrm{v30}}}}%
\def\Ebar{\ensuremath{E\hspace{-.23cm}/\hspace{+.01cm}}}
\def\EM25{\ensuremath{\Ebar_{25}}}
\def\EMF25{\ensuremath{\Ebar^{\perp}_{25}}}
\def\ECM60{\ensuremath{\Ebar^b_{60}}}

\def\TKM25{\ensuremath{N_{tk}^{25}}}

\def\DELTAM{\ensuremath{\Delta{M}}}%

\def\simge{\ \lower -2.5pt\hbox{$>$} \hskip-8pt \lower 2.5pt \hbox{$\sim$}\ }
\def\simle{\ \lower -2.5pt\hbox{$<$} \hskip-8pt \lower 2.5pt \hbox{$\sim$}\ }

\newlength{\capindent}
\newlength{\capwidth}
\newlength{\figwidth}
\newcommand{\icaption}[2][!*!,!]{\hspace*{\capindent}%
  \begin{minipage}{\capwidth}
    \ifthenelse{\equal{#1}{!*!,!}}%
      {\caption{#2}}%
      {\caption[#1]{#2}}
  \end{minipage}}
\begin{document}
\begin{titlepage}
\title{Search for Scalar Leptons
\boldmath in e$^+$e$^-$ collisions  \\ at $\sqrt{s}=189 \gev$}
%\author{A. Favara, S. Rosier-Lees}
\author{The L3 collaboration}
\vspace{1.5cm}

%
% The abstract
%
\begin{abstract}
We report the result of a search for scalar leptons in \epem\, 
collisions at 189 \gev{} centre-of-mass energy at LEP.
No evidence for such particles is found in a data sample of 176 pb$^{-1}$.
Improved upper limits are set on the production cross sections for 
these new particles.
New exclusion contours in the parameter space of the
Minimal Supersymmetric Standard Model are derived, 
as well as new lower limits on the masses of these
supersymmetric particles.
Under the assumptions of common gaugino and scalar masses at the GUT
scale, we set an absolute lower limit on the mass of the lightest
scalar electron of 65.5 \gev.
\end{abstract}

\submitted
%\vspace*{20mm}
%\centerline{Submitted to 
%             {\it International Europhysics Conference High Energy
%               Physics 99 }}

%\centerline{Tampere, Finland, 15-21 July 1999.}

\end{titlepage}

%
%

%%%%%%%%%%%%%%%%%%%%%%%%%%%%%%%%%%%%%%%%%%%%%%%%%%%%%%%%%%%%%%%%%%%%%%%%%%%%%%%
% Introduction
%%%%%%%%%%%%%%%%%%%%%%%%%%%%%%%%%%%%%%%%%%%%%%%%%%%%%%%%%%%%%%%%%%%%%%%%%%%%%%%
%
\section{Introduction}

One of the main goals of the LEP experiments is to search for new
particles predicted by theories beyond the Standard
Model. In this letter we report on searches for 
unstable scalar leptons.
These particles are predicted by supersymmetric theories (SUSY)
\cite{susy}. In SUSY theories with minimal particle
content (MSSM) \cite{mssm}, in addition to the ordinary particles, there is
a supersymmetric spectrum of particles with spins which differ by one
half with respect to their Standard Model partners.

Scalar leptons ($\slepton{\pm}_R$ and $\slepton{\pm}_L$) are the 
supersymmetric partners of the right- and left-handed leptons.
Pair production takes place through $s$-channel $\gamma/\mathrm{Z}$
exchange. For scalar electrons the production cross section 
is enhanced by $t$-channel exchange of a neutralino.

Short-lived supersymmetric particles are expected
in R-parity conserving SUSY models. 
The R-parity is a quantum number which 
distinguishes ordinary particles from supersymmetric particles.
If R-parity is conserved supersymmetric particles are 
pair-produced and the lightest supersymmetric particle, the lightest neutralino
\neutralino{1}, is stable. The neutralino is weakly-interacting 
and escapes detection.
In this letter we assume R-parity conservation, which implies 
that the decay chain of supersymmetric particles always contains,
besides standard particles, two invisible neutralinos
causing the missing energy signature.

The scalar lepton decays into its partner lepton mainly via
$\slepton{\pm} \rightarrow \neutralino{1} \ell^\pm$, but also via the cascade
decay, such as $\slepton{\pm} \rightarrow \neutralino{2} \ell^\pm
\ra \neutralino{1}\Zzv \ell^\pm $, which may dominate in some regions
of the parameter space of the MSSM. 
%In general, the SUSY partners of 
%the right-handed leptons ($\slepton{\pm}_R$)
%are expected to be lighter than their counterparts for left-handed leptons.

Previous limits on scalar leptons have been obtained 
at lower energies by L3~\cite{reflep1.5,susy_96,nota99slep} 
and other LEP experiments~\cite{lepsle98}.
Results presented in this paper are organised as follows:
Data sample and event simulation are presented in Section 2;
Experimental signatures and event selections are discussed in Section
3; In Section 4
 experimental results are summarised and in Section  5 model
independent limits are presented on production cross sections.
In Section 6, our experimental results are interpreted in 
the framework of the constrained MSSM, and in the minimal
supergravity model. In these models, lower limits on the masses of 
supersymmetric particles are obtained.
For these limits present experimental results are combined 
with those obtained previously by L3 at the Z peak \cite{oldsusyl3} 
and at energies up to 183 \gev{} 
\cite{reflep1.5,susy_96,nota99slep}.

\section{Data Sample and Simulation} \label{dtsmcs}

We present the analysis of data collected with the L3
detector \cite{l3-detector} in 1998, 
corresponding to an integrated luminosity of 176.3 pb$^{-1}$ at an
average centre-of-mass energy, $\sqrt{s}$, 
 of 188.6 \gev, denoted hereafter as $\sqrt{s}=189 \gev$.

Standard Model reactions are simulated with the following 
Monte Carlo generators:
{\tt PYTHIA}~\cite{PYTHIA} for 
  $\ee \rightarrow \mathrm{q\bar{q}}$,
  $\ee \rightarrow \mathrm{Z}\,\ee$ and 
  $\ee \rightarrow \gamma\!/\mathrm{Z}\,\gamma\!/\mathrm{Z} $;
{\tt EXCALIBUR}~\cite{EXCALIBUR} for
   $\ee \rightarrow \mathrm{W^\pm\, e^\mp \nu}$;
{\tt KORALZ}~\cite{KORALZ} for
   $\ee \rightarrow \mu^+\mu^-$ and
   $\ee \rightarrow \tau^+\tau^-$;
{\tt BHWIDE}~\cite{BHWIDE} for 
   $\ee \rightarrow \ee$;
{\tt KORALW}~\cite{KORALW} for
   $\ee \rightarrow \mathrm{W^+ W^-}$;
%%% {\tt GGG}~\cite{GGG} for
%%%   $\ee \rightarrow \gamma\gamma$. 
two-photon interaction processes have been simulated using 
{\tt DIAG36}~\cite{DIAG} ($\ee \rightarrow \ee \ell^+\ell^-$) and
{\tt PHOJET}~\cite{PHOJET} ($\ee \rightarrow \ee\, \mathrm{hadrons}$), 
requiring at least 3 \gev{} for the invariant mass of the two-photon system.
The number of simulated events for each background process is 
equivalent to more than 100 times the statistics of the collected 
data sample except for two-photon interactions for which it is more 
than two times the data statistics.

Signal events are generated with the Monte Carlo program 
{\tt SUSYGEN}~\cite{susygen2.2}, for masses of SUSY particles 
($M_{\rm SUSY}$) ranging from 45 \gev{} up to the kinematic limit, 
 and for $\DELTAM$ values 
($\dm = M_{\rm SUSY} - M_{\neutralino{1}}$) between 3 \gev{} and
$M_{\rm SUSY}-1 \GeV$. 

The detector response is simulated using the {\tt GEANT} 
package~\cite{geant}. It takes into account effects of energy loss,
multiple scattering and showering in the detector materials and
in the beam pipe. Hadronic interactions are simulated with the
{\tt GHEISHA} program~\cite{gheisha}. Time dependent inefficiencies
of the different subdetectors are also taken into account
in the simulation procedure.

%
%%%%%%%%%%%%%%%%%%%%%%%%%%%%%%%%%%%%%%%%%%%%%%%%%%%%%%%%%%%%%%%%%%%%%%%%%%%
% Experimental Strategy
%%%%%%%%%%%%%%%%%%%%%%%%%%%%%%%%%%%%%%%%%%%%%%%%%%%%%%%%%%%%%%%%%%%%%%%%%%%
%
\section{Analysis Procedure}

\subsection{Signal topologies and optimisation procedure}
\label{sec:optimization}

Besides the main characteristic of missing transverse momentum,
supersymmetric particle signals can be further specified according
to the number of leptons or the multiplicity of hadronic jets in the
final state. 

For scalar leptons, signatures are simple since most of the time
the final state is given by two acoplanar leptons of the same generation.
To account for the three lepton types three different selections are
performed. While for scalar electrons and muons, two identified
leptons are required in the event, scalar taus are selected as low 
multiplicity events with two narrow jets.

A new analysis searching for a single electron is also presented for
the first time. This search accounts for 
$ \ee \rightarrow \susy{e}_R\susy{e}_L$
production when the $\susy{e}_R$ is almost mass degenerate with the
$\neutralino{1}$, since the $\susy{e}_L$ is heavier than the $\susy{e}_R$.
Thus the visible electron comes from the decay of
$\susy{e}_L\ra\neutralino{1}\rm e$, while the decay of $\susy{e}_R$
can be invisible for $\dm \sim 0$.

The \dm\, dependence of the signal and background events is taken
into account with different optimisations for each selection. For scalar 
electron and scalar muon analyses, three selections are performed: for 
the low \dm\, range at $3-5 \gev$, 
the medium \dm\, range at $10-40 \gev$ and the high
\dm\, range at $60-90 \gev$. For the scalar tau analysis, 
four selections are optimised for different \dm\ ranges: 
$3-7 \gev$, $7-15 \gev$, $15-30 \gev$, above $30 \gev$.

The cut values of each selection are {\it a priori} optimised using
Monte Carlo signal and background events. 
The optimisation procedure varies all cuts simultaneously to maximise the
signal efficiency and the background rejection.
In fact, we minimise the average limit ($\kappa^{-1}$), 
for infinite number of experiments, 
assuming only background contribution. This is expressed mathematically
by the following formula:
\begin{equation}  
\kappa=\epsilon/ \Sigma_{n=0}^{\infty} k(b)_{n} P(b,n)
\end{equation}
where $k(b)_{n}$ is the 95\% confidence level Bayesian upper limit, 
$P(b,n)$ is the Poisson distribution 
for $n$ events with an expected background of $b$ events, and 
$\epsilon$ is the signal efficiency.

\subsection{Event selection}

Lepton and photon identification, and isolation criteria in hadronic events
are unchanged compared to our previous analysis at $\sqrt s=183$ \gev{}
\cite{susy_96}.
The Durham algorithm~\cite{durham} is used for the clustering of
hadronic jets.

Events are first selected by requiring at least 3 \gev{} of visible energy
and 3 \gev{} of transverse momentum. Beam-gas events are rejected by
requiring the visible energy in a cone of $30^\circ$  around the beam
pipe to be less than 90\% of the total, and the missing momentum
vector to be at least $10^\circ$ away from the beam pipe.
Tagged two-photon interactions are rejected by requiring the sum of
the energies measured in the lead-scintillator
ring calorimeter and in the luminosity monitors~\cite{l3-detector}
to be less than 10 \GeV.
These two detectors cover the polar angle range 
$1.5^\circ < \theta <9^\circ$ on both sides of the interaction 
point.

Given the low multiplicity of the
signal, events are rejected if the number of tracks is larger than 6 or if the
number of calorimetric clusters ($N_{cl}$) is larger than 15.
We then require two or three identified leptons and photons.
The following quantities are defined:
the energy depositions
($E^{\perp}_{25}$ and $E_{25}$)
within $\pm25^\circ$ around the missing 
energy direction in the R--$\phi$ plane 
 or in space respectively, and the energy deposition in a $60^\circ$ half opening
angle cone around the vector opposite 
to the sum of the two jet  
directions in space ($E_{60}^{b}$). 
% When three leptons are identified, a cut on the
%sum of the three angles  between the three lepton pairs ($\TRE$) is also
%applied. 
We also apply
cuts on the lepton energy ($E_{lep}$),
on the total transverse momentum of the leptons ($p_{\perp}$), on their maximum 
acollinearity and acoplanarity, on the 
polar angle of the missing energy vector ($\theta_{miss}$) and 
on the variable $E_{TTL}$.
The latter is defined as the absolute
value of the projection of the total momentum of the two highest
energy leptons onto the
direction perpendicular to the leptonic thrust computed in the R-$\phi$ 
plane. 

\begin{figure}[hbtp]
\begin{center}
\psfig{file=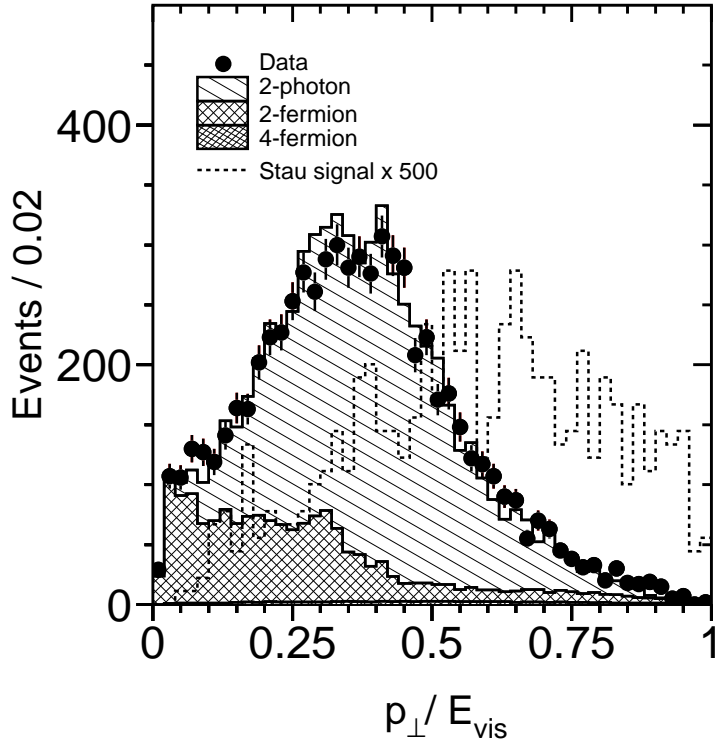,width=11.0cm}
%\mbox{\epsfxsize=10.0cm \epsffile{stau_189.eps} }
  \caption{Normalised transverse momentum
distributions  $p_{\perp}/E_{vis}$  for
data and  MC events at $\sqrt{s}=189 \GeV$ after
preselection. Contributions from 2-photon interactions, 2-fermion 
and 4-fermion final states are given separately. The 
distribution for an expected scalar tau signal  
with $M_{\tilde{\tau}_R}=70 \GeV{}$ and $M_{\tilde{\chi}_1^0}=55 \GeV{}$
is also shown.\label{fig:staubis}}
\end{center}
%\end{figure}

% 
%
%\vspace{-1 cm}
%\begin{figure}[hbtp]
\end{figure}

The scalar taus are selected as
low multiplicity events with acoplanar jets.
Upper cuts on the jet width $y_{\bot}$, defined as
the ratio between the sum of particle momenta 
transverse to the jet direction and the jet energy, are also applied. 
Distributions of the normalised transverse
missing momentum $p_{\perp}/E_{vis}$ 
are shown in Figure \ref{fig:staubis} 
for data, signal and background
Monte Carlo events, at the preselection level.

The cut values optimised 
at $\sqrt{s}=189 \gev$ for the scalar lepton searches
are quoted in Table~\ref{tab1} for the different \dm\, ranges. 

The single electron analysis makes use of very simple requirements
aimed at a reliable identification of the electron and a nearly
empty detector elsewhere.
If two tracks are detected, their acoplanarity must be between
$10^\circ$ and $160^\circ$. The electron energy has to be less
than 65 \gev{} to reject photon conversion from 
$\ee \rightarrow \nu\bar\nu\gam$, when
the two tracks are not resolved.
The energy of a second electron should be less
than 4 \gev, and its acoplanarity with respect to the highest
energy electron must be at least $20^\circ$.
If only one electron is detected, the missing transverse momentum is
required to be at least 6 \gev. If a second electron of at least
100 \mev{} is detected, the missing transverse momentum must be greater
than 10 \gev.

\section{Results}

The results obtained at $\sqrt{s}=189 \gev$ 
for the ten scalar lepton selections
are shown in Table \ref{tab8}. In this table, the results for 
the two lowest \dm\, selections for scalar taus are shown together.
A good agreement between the expected background from Standard Model
processes and the selected data is observed.

The ten scalar lepton analyses performed at $\sqrt{s}=189 \gev{}$ select 
21, 19 and 56 candidates in 
 the scalar electron, muon and tau channels, respectively. 
Those observations are in good
agreement with the background expectation of 23.0, 21.0 and 51.6 events,
 respectively.

\begin{figure}
\begin{tabular}{cc}
\psfig{file=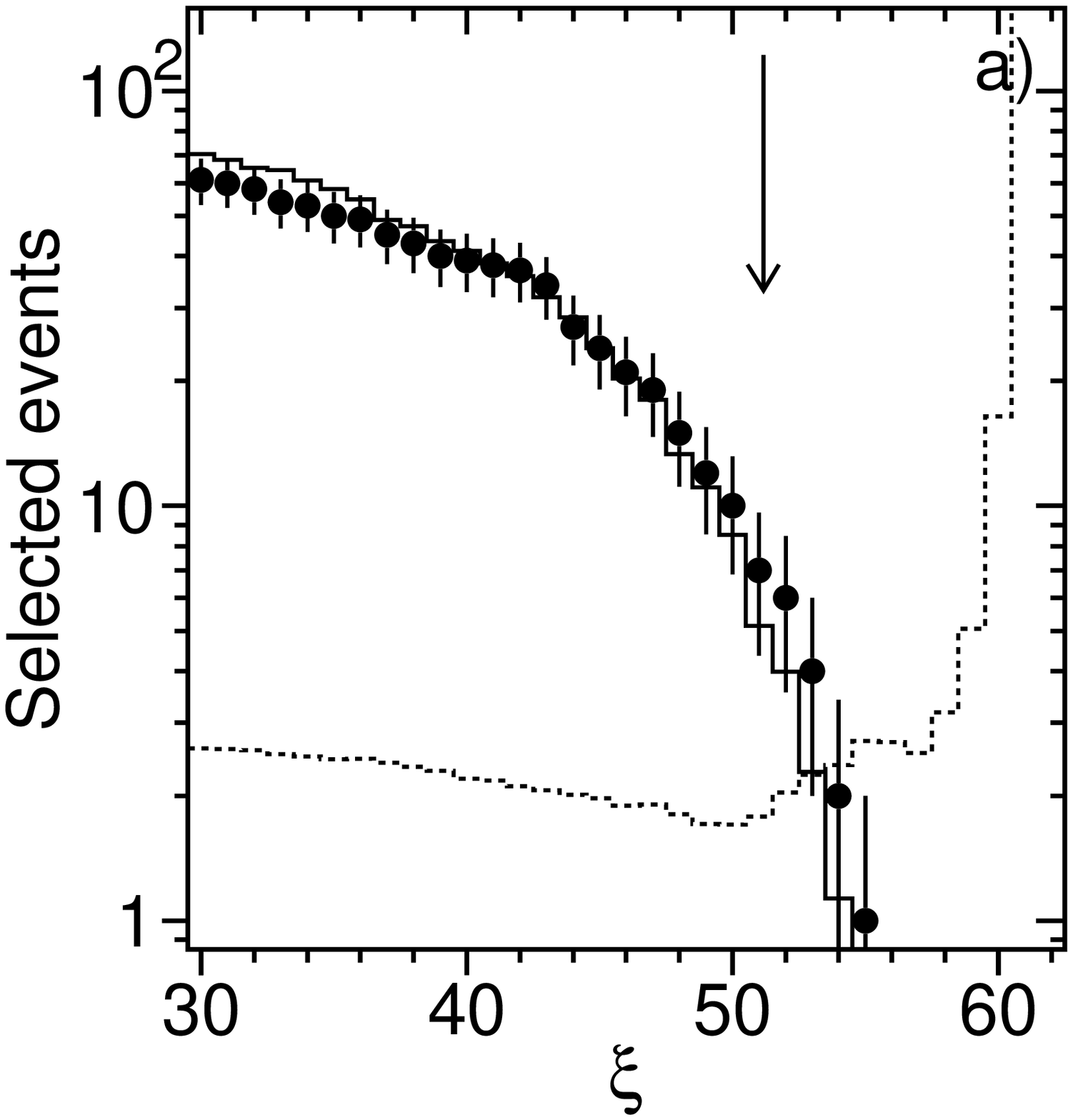,width=9.0cm} &
\psfig{file=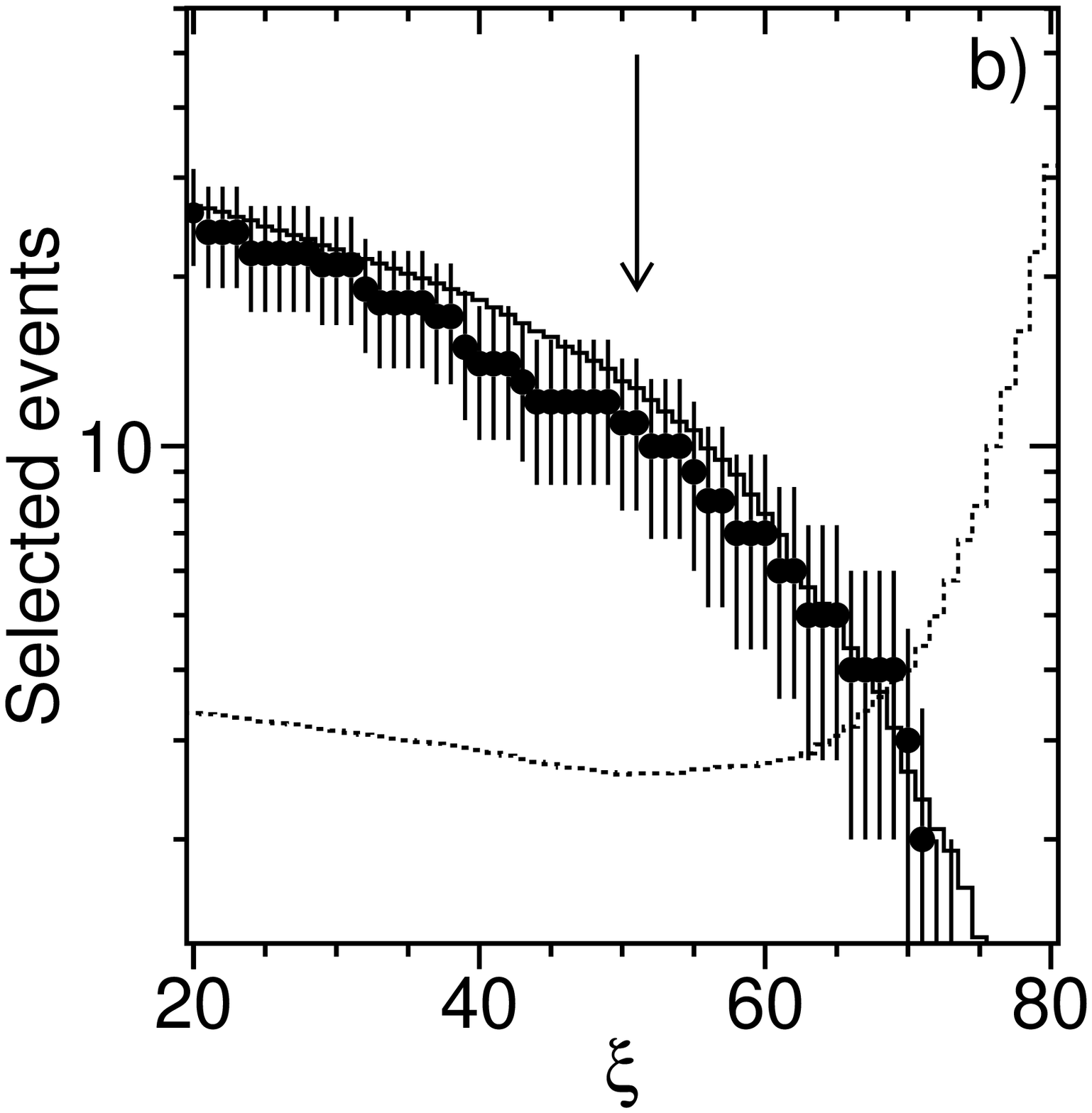,width=9.0cm}  \\
\psfig{file=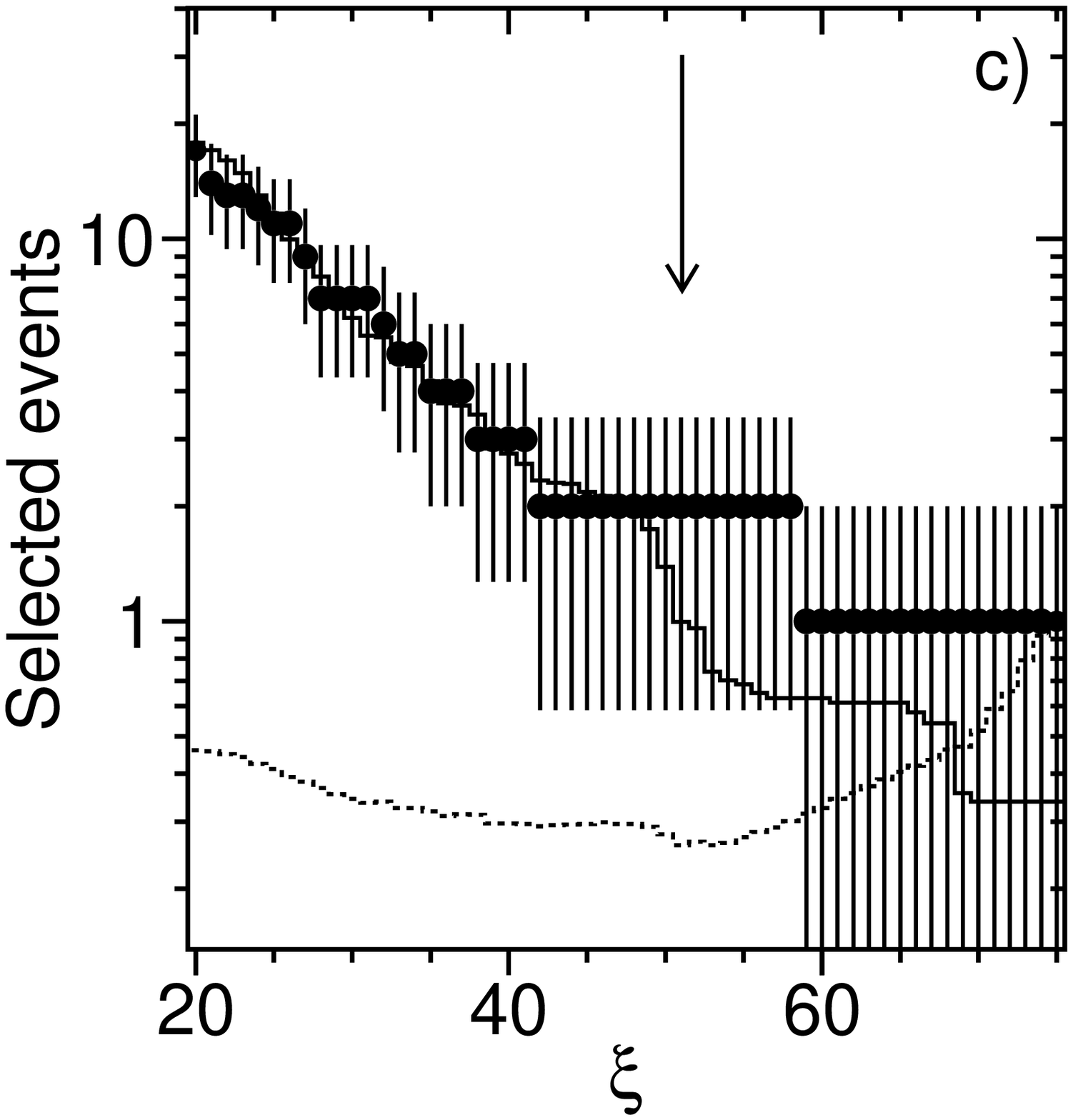,width=9.0cm} &
\psfig{file=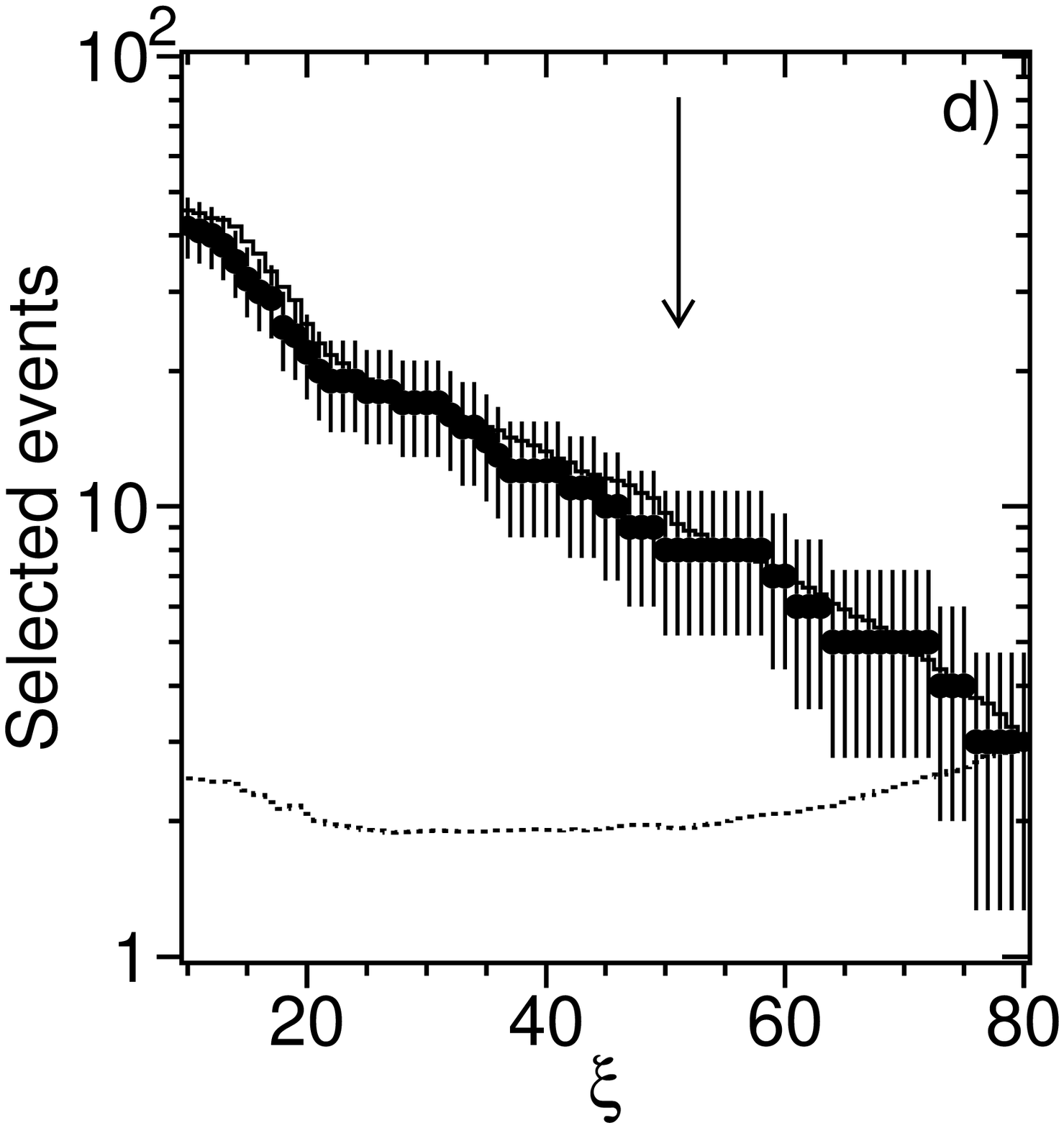,width=9.0cm}  \\
\end{tabular}
\caption{Number of events selected in data (dots),
in Monte Carlo simulations of standard processes (solid line)
and signal sensitivity (dashed line)
as a function of selection cuts with increasing background rejection power.
The vertical arrows show the $\xi$ value corresponding to
the optimised cuts.
Distributions for the scalar electron low \dm\, a) and high \dm\,
b), the scalar muon medium \dm\, c) and high \dm\, d) selections
are shown. \label{fig:xi_sleptons}}
\end{figure}

All the scalar lepton selections are parametrised
as a function of a single parameter, $\xi$,
in the following manner: 
given a lower edge, $X_{loose}^i$, and an upper edge, $X_{tight}^i$,
for the cut on variable $i$, the parameter $\xi$ is equal to 0
when this cut is at the lower edge (many background events satisfy
the selection) and to 100 when it is at the upper
edge (no or few background events pass the selection). 
All cuts ($i=1,....,N$) are related to the parameter $\xi$ as follows:
$$X_{cut}^i=X_{loose}^i + {(X_{tight}^i- X_{loose}^i)} \times \frac{\xi}{100}.
$$
The parameter $\xi$ is scanned around the optimal value ($\xi=50$)
to check the agreement
between data and Monte Carlo at different background rejection stages.
As illustrated in Figure~\ref{fig:xi_sleptons} 
for electrons and muons, and for several \dm\, ranges,
the data and Monte Carlo simulations are in good agreement.
The vertical arrows show the $\xi$ value corresponding to
the optimised cuts.

For intermediate \dm\, values different from those chosen for
optimisation, we choose the combination of selections among those 
previously  defined, providing the highest sensitivity \cite{susy_96}.
In this combination procedure, we take into account the overlap
among the selections within the 
data and Monte Carlo samples.

The selection efficiencies
at $\sqrt{s}=189 \gev$ for scalar lepton pair production, as well as the background
expectations, are reported for different values of \dm\, in Table~\ref{tab5}.
Efficiencies vary from 19\% to 58\% for scalar electrons and
from 11\% to 36\% for scalar muons.
In comparison, the scalar tau selection efficiencies are smaller, ranging 
from 1.4\% to 30\%.        

\begin{figure}
\begin{center}
\psfig{file=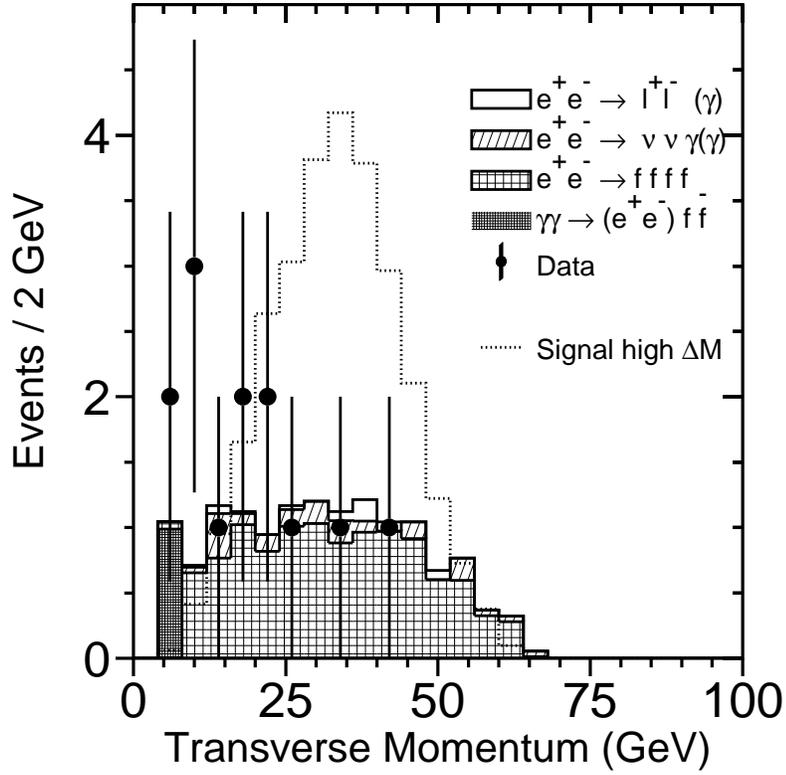,width=12.0cm}
%  \mbox{\epsfxsize=10.0cm \epsffile{prsingle98.eps}}
  \caption{Transverse momentum distribution for the selected events
 in the single electron final state analysis.
 Data events observed at $\sqrt{s}=189 \gev$
are compared to Standard Model background processes 
 and to the expected signal distributions with arbitrary 
normalisation.
\label{fig:datamcse}}
\end{center}
\end{figure}

With the single electron analysis,
13 events are selected in data and 14.0 are
expected from Standard Model processes.
The transverse momentum distributions 
for the selected data, signal and background Monte Carlo events
are shown in Figure \ref{fig:datamcse}.
Signal efficiencies vary from 4\% at $m_{\susy{e}_L}-m_{\neutralino{1}}=5 \gev$ to 60\%
at $m_{\susy{e}_L}-m_{\neutralino{1}}=50 \gev$, and they do not change
for any values of $m_{\susy{e}_R}-m_{\neutralino{1}}$ smaller than 4 \gev.

Systematic errors on the signal efficiencies are evaluated as in
Reference \citen{reflep1.5}, and they are typically 5\% relative, 
dominated by Monte Carlo statistics. These errors
are taken into account following the procedure explained 
in Reference \citen{cal_limit}.

\section{Model independent upper limits on production cross sections}
No excess of events is observed
and we set upper limits on scalar lepton
 production cross sections.
%as well as limits on the masses of these
%particles in the framework of the MSSM.
Exclusion limits at 95\% C.L. are derived taking into account background
contributions.

To derive the new upper limits on the production cross sections,
and for interpretations in the MSSM we combine the 1998 data sample
collected at $\sqrt{s}=189 \gev$
 with those collected at lower centre-of-mass energies.

\begin{figure}
\begin{tabular}{cc}
\psfig{file=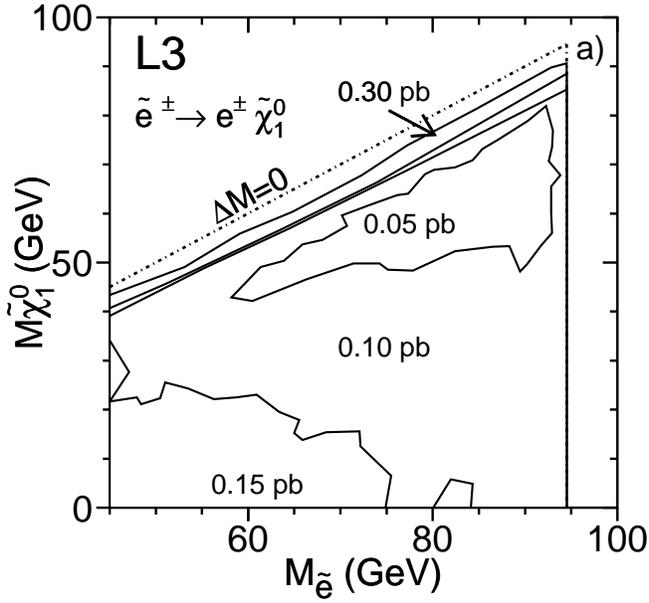,width=9.0cm} &
\psfig{file=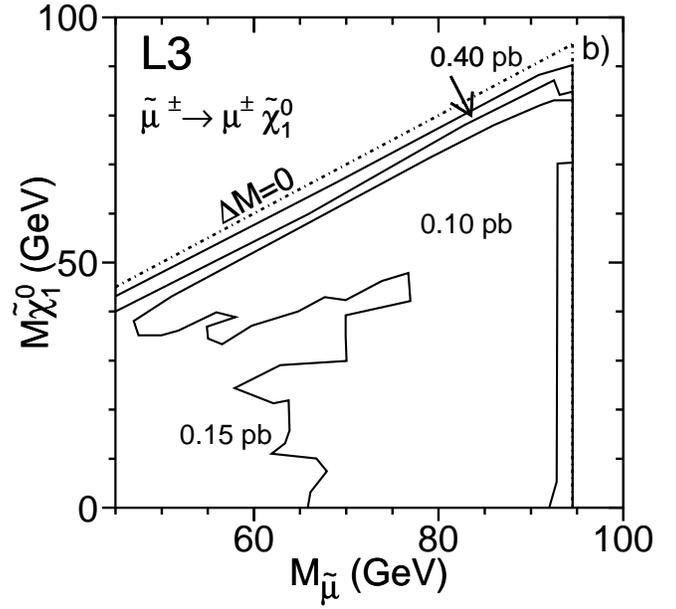,width=9.0cm}
  \\
 \psfig{file=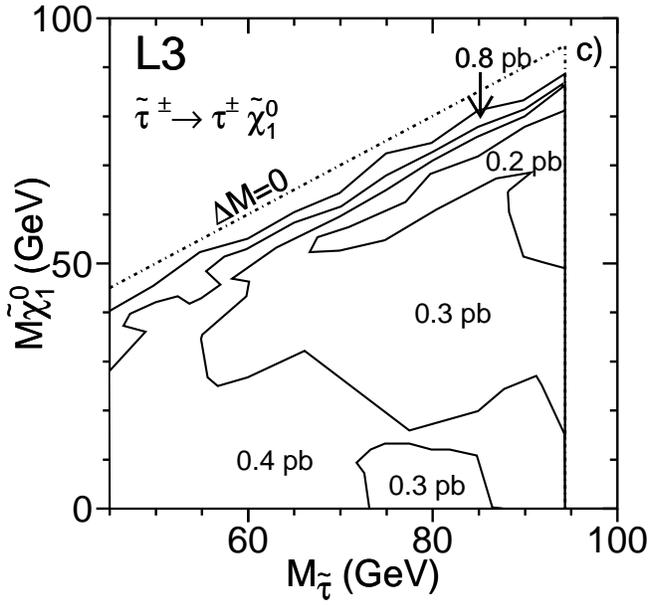,width=9.cm} &
  \\
\end{tabular}
 \caption{Upper limits on the production cross sections up 
to  $\sqrt{s}=189 \gev$ shown in the mass plane 
$M_{\susy{\ell}}-M_{\neutralino{1}}$
   for scalar leptons.       Figures  a), b) and c)
 show the limits for scalar electrons, muons and  taus, respectively.
 \label{fig:xsection_sleptons}}
\end{figure}

Assuming a branching fraction for
$\slepton{\pm} \rightarrow \neutralino{1} \ell^\pm$ of 100\%, upper
limits are set on pair production cross sections of scalar electrons,
muons and taus in the plane 
$M_{\neutralino{1}}$ versus $M_{\slepton{\pm}}$ as depicted in 
Figure~\ref{fig:xsection_sleptons}.
The efficiency for the selection of scalar electrons includes the $t$-channel
contribution.
For scalar electron and muon masses below $94 \gev$, and \dm\ sufficiently
large, cross sections above $0.1$~pb  are excluded.
Owing to the lower selection sensitivity, the corresponding upper
limit for the scalar tau cross section
is $0.3$~pb.

\section{Limits on scalar lepton masses in the MSSM}

In the MSSM, with Grand Unification assumptions~\cite{MSSM_GUT},
the masses and couplings of the SUSY
particles as well as their production cross sections,
are entirely described~\cite{mssm} once five parameters are fixed:
$\tan\beta$, the ratio of the vacuum expectation values of the two Higgs 
doublets, $M \equiv M_2$, the gaugino mass parameter,
$\mu$, the higgsino mixing parameter,
$m_0$,  the common mass for scalar fermions at the GUT scale, and $A$, 
the trilinear coupling in the Higgs sector.
We investigate the following MSSM parameter space: 
$$
   \begin{array}{rclcrcl}
         1 \leq& \tan\beta & \leq 60 ,     &&
0 \gev \leq &M_2 &\leq 2000 \gev ,\\
 -2000 \gev \leq& \mu       & \leq 2000 \gev,&& 
0 \gev \leq& m_0         &\leq  500 \gev.
  \end{array}
$$

All the limits on the cross sections previously shown combined with
the results obtained at lower centre-of-mass energies, and
for the mSUGRA interpretation with the recent results of chargino and 
neutralino searches \cite{charg99ref}, 
can be translated into exclusion regions in the MSSM parameter space.
To derive limits in the MSSM, 
we optimise the global selection for any different
point in the parameter space. This is obtained, choosing every time the
combination of selections providing the highest sensitivity, given 
the production cross sections and the decay branching fractions which are 
calculated with the generator {\tt SUSYGEN}.

\begin{figure}[hbtp]
\begin{tabular}{cc}
\psfig{file=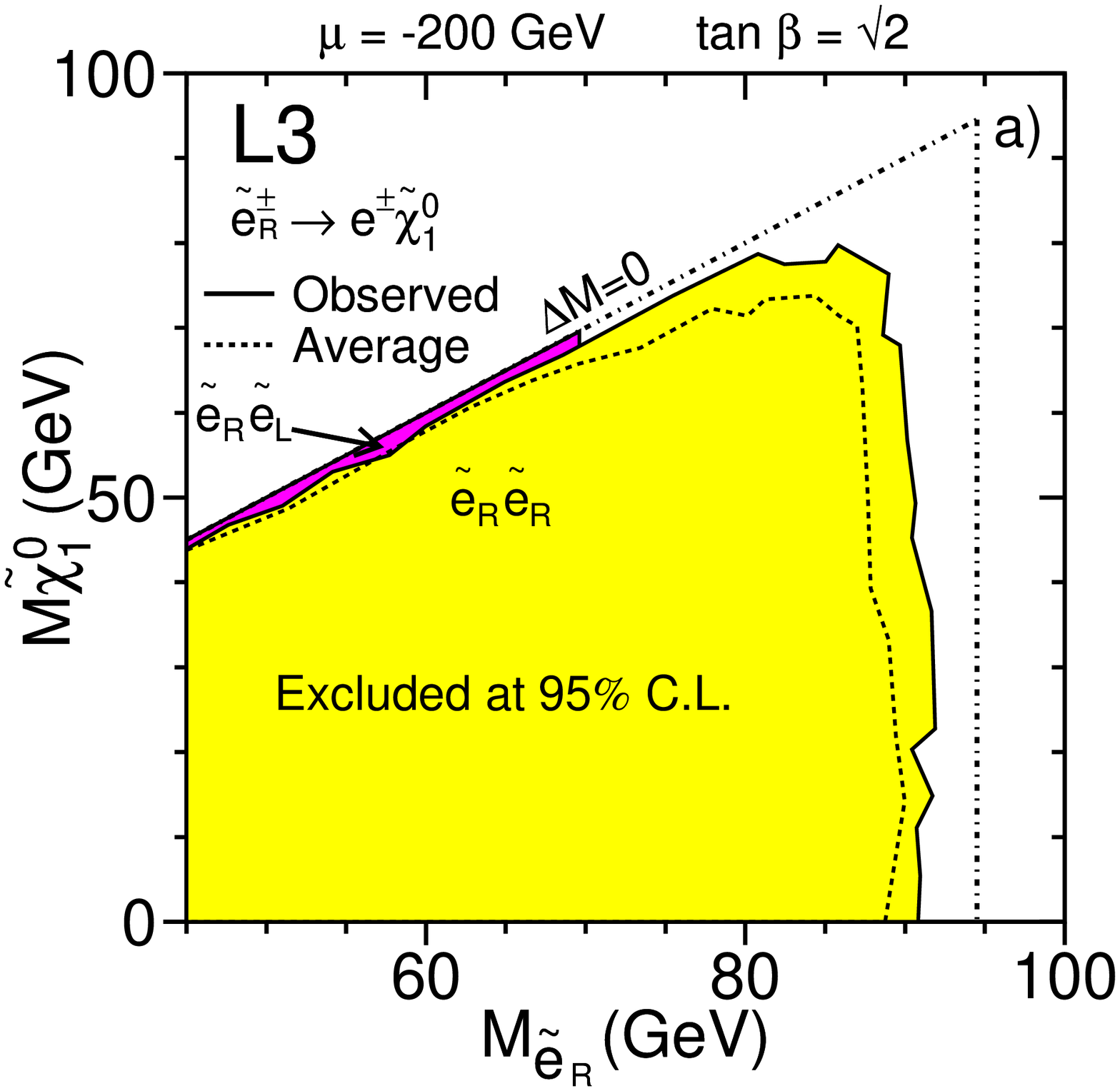,width=9.cm} &
\psfig{file=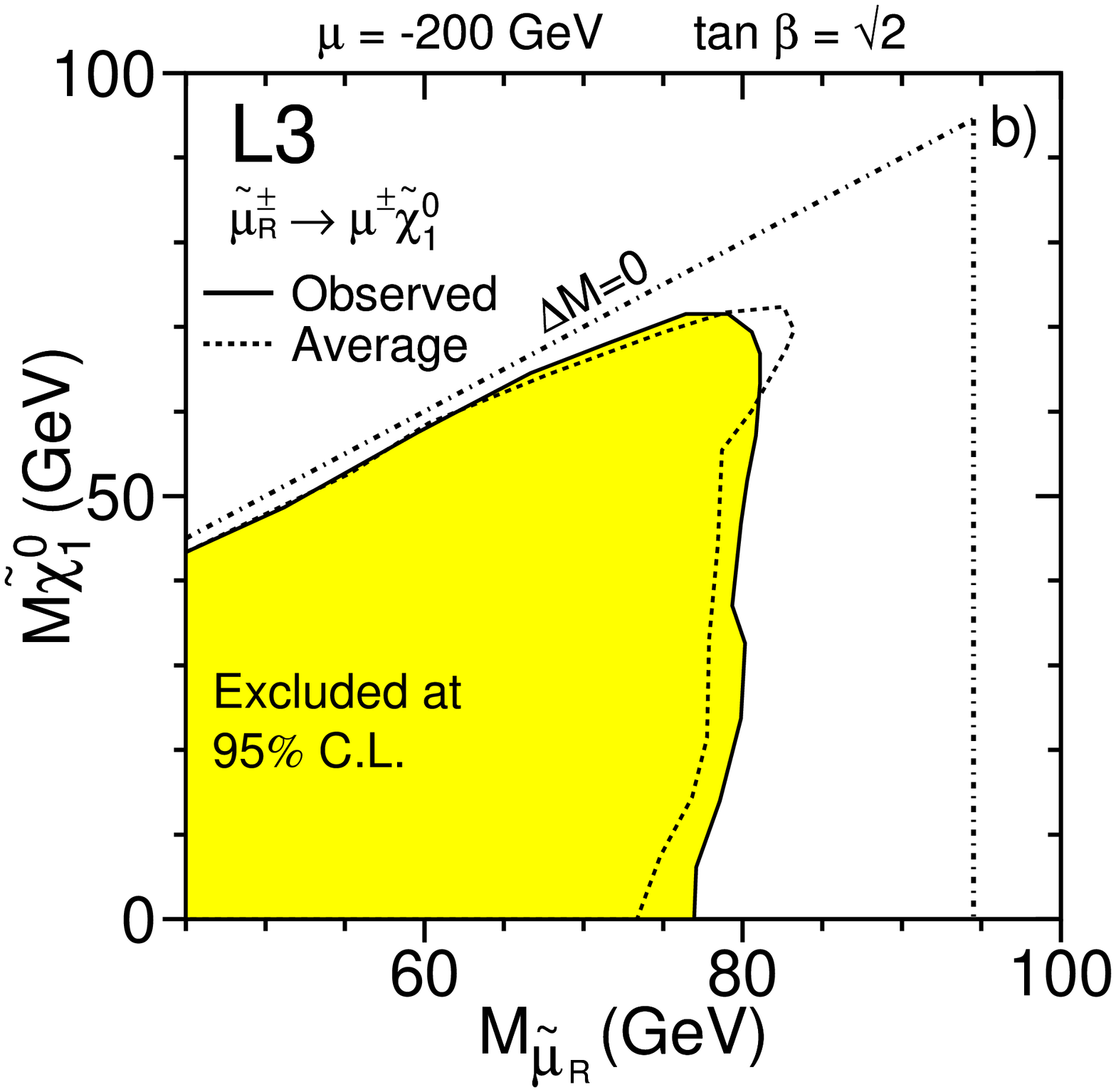,width=9.cm} 
\\  
 \psfig{file=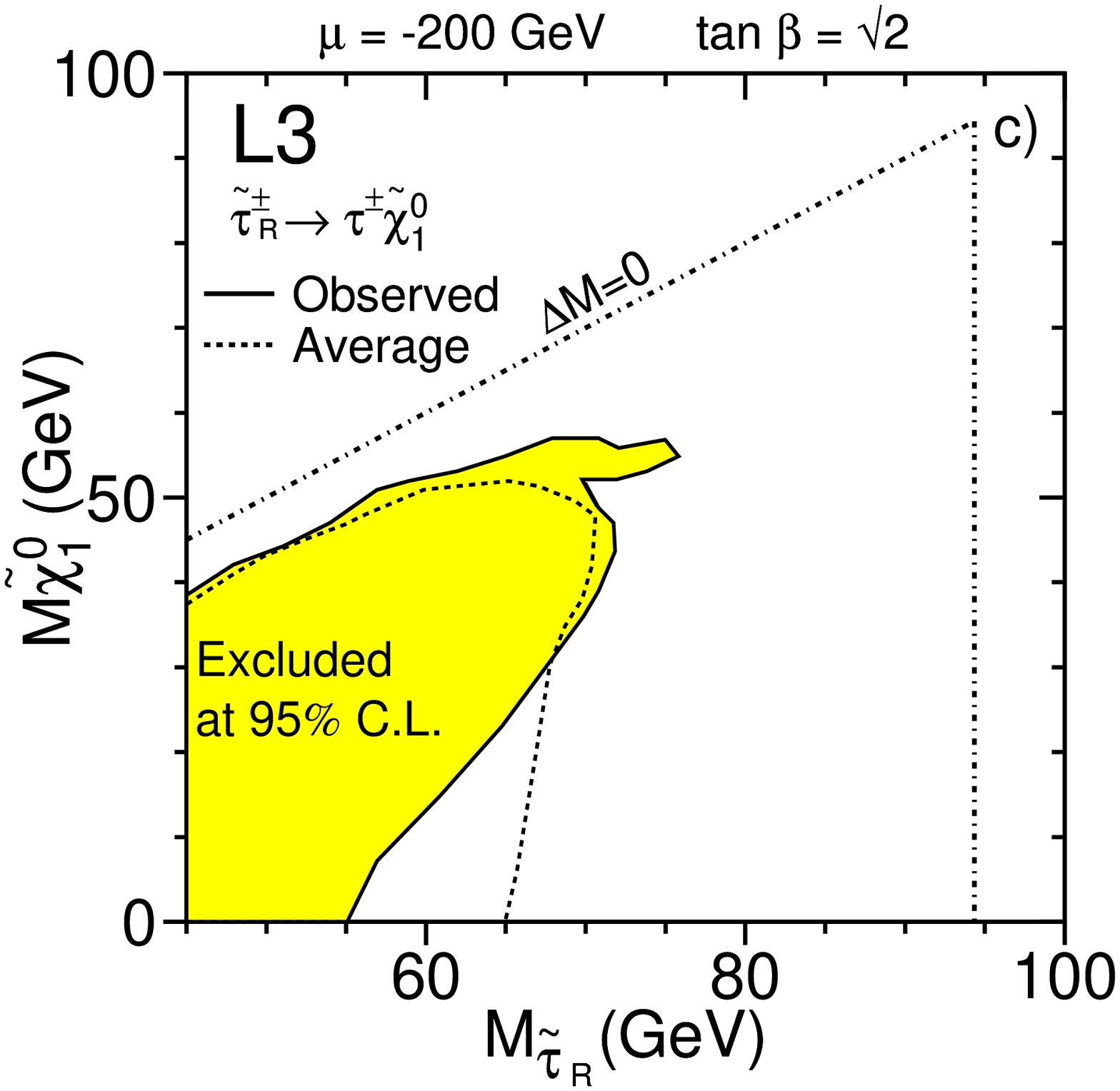,width=9.cm} &  
\psfig{file=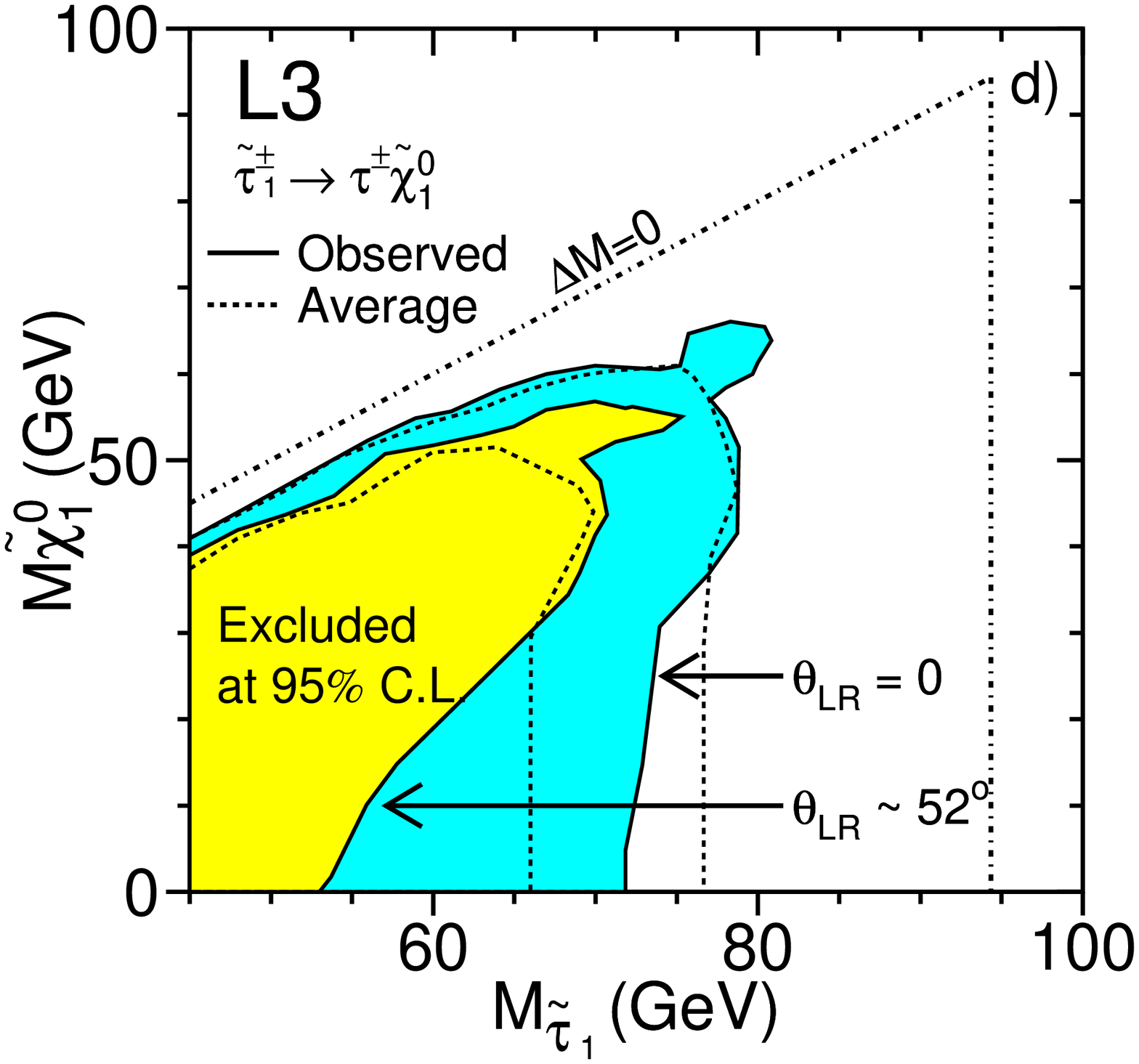,width=9.cm}\\
\end{tabular}
 \caption{Mass limits on the scalar partners of right-handed  
electrons a), muons b)  and taus c) as a function of the neutralino mass
          $M_{\neutralino{1}}$. d) shows the exclusion for the
          scalar tau, when mixing between $\susy{\tau}_R$ and $\susy{\tau}_L$
          occurs, for the minimal and maximal cross sections.
          These four figures are obtained using only the upper limits on the
          cross section from direct searches at
          centre-of-mass energies between 130 \gev{} and 189 \gev. The
          dashed lines show the average limits obtained with Monte
          Carlo trials with background only.
            \label{fig:limit_rleptons}}
\end{figure}

In general, the SUSY partners of the right-handed leptons ($\slepton{\pm}_R$)
are expected to be lighter than their counterparts for left-handed leptons.
Hence, we show in Figures \ref{fig:limit_rleptons}a, \ref{fig:limit_rleptons}b
and \ref{fig:limit_rleptons}c the exclusion contours 
in the $M_{\neutralino{1}} - M_{\slepton{\pm}_R}$ plane
considering only the reaction
$\ee \rightarrow \slepton{\pm}_R \, \slepton{\mp}_R$ and setting
$\mu = -200 \gev$ and $\tan\beta = \sqrt{2}$. These exclusions
hold also for higher $\tan\beta$ and $|\mu|$ values. For smaller $|\mu|$
values, the $t$-channel contribution to the scalar electron cross
section is reduced, thus reducing by a few \gev{} the limit on its
mass shown in Figure~\ref{fig:limit_rleptons}a.
The values of $\mu$ and $\tan\beta$ are also relevant 
for the calculation of the branching ratio for the decay
$\slepton{\pm} \rightarrow \neutralino{2} \ell^\pm\ra\neutralino{1}\Zzv \ell^\pm$ 
in Figures \ref{fig:limit_rleptons}a--c.
To derive these exclusions, only the purely leptonic decay
$\slepton{\pm}_R \rightarrow \ell^\pm\neutralino{1}$ is considered,
neglecting any additional efficiency from cascade decays.

Under these assumptions lower limits on scalar lepton masses
are derived. From Figures \ref{fig:limit_rleptons}a and \ref{fig:limit_rleptons}b
scalar electrons lighter
than 85.5 \gev, for $\dm > 10 \gev$, and scalar muons lighter than 78 \gev,
for $\dm > 15 \gev$, are excluded. 
Including also the contribution from the process
$\epem\ra\susy{e}_R\susy{e}_L$ and using the single electron
selection, the very small \dm{} region for the $\susy{e}_R$ can be excluded
at 95\% C.L. up to $M_{\selectron{\pm}_R}=69.6 \gev$. This additional
exclusion is shown as the dark area in Figure~\ref{fig:limit_rleptons}a.
From Figure
\ref{fig:limit_rleptons}c we conclude that scalar taus lighter than 65 \gev,
for $10 \gev < \dm < 40 \gev$, are excluded if there is no mixing.

Mass eigenstates of scalar leptons are in general a mixture of
the weak eigenstates $\slepton{\pm}_R$ and $\slepton{\pm}_L$.
The mixing between $\slepton{\pm}_R$ and $\slepton{\pm}_L$
is proportional to the mass of the partner lepton.
Hence the mixing for scalar electrons and muons is always negligible
while it can be sizable for scalar taus.
The mixing is governed by the parameters $A$, $\mu$ and $\tan\beta$.

Scalar tau mass eigenstates are given by
$ \stau{}_{1,2} =  \stau{}_{L,R} \cos\theta_{\mathrm{LR}} \pm
\stau{}_{R,L}\sin\theta_{\mathrm{LR}}$,
where $\theta_{\mathrm{LR}}$ is the mixing angle.
The production cross section for scalar taus can be parametrised
as a function of the scalar tau mass and of the mixing angle
\cite{stopstau}.
At $\theta_{\mathrm{LR}}\sim 52^\circ$ the scalar tau
decouples from the Z and the cross section is minimal.
It reaches the maximum at $\cos\theta_{\mathrm{LR}}$=1 when
the scalar tau is equivalent to the weak eigenstate $\stau{\pm}_L$.

The exclusion contours in Figure~\ref{fig:limit_rleptons}d are obtained
considering only the reaction $\ee \rightarrow \stau{+}_1\,\stau{-}_1$ 
and assuming 100\% branching ratio for $\stau{}_{1} \rightarrow \tau
\tilde{\chi}_1^0$. The two
contours correspond to the minimal and maximal cross sections.
Under the most conservative assumption for the mixing, a
scalar tau lighter than 60 \GeV{} is excluded for 
$\dm$ values between 8 and 42 \GeV. In case of
$\cos\theta_{\mathrm{LR}}=1$ the mass limit reaches
71.5 \GeV{} assuming $\dm$ greater than 12 \GeV.

\begin{figure}
\begin{center}
  \mbox{\epsfxsize=9.5cm \epsffile{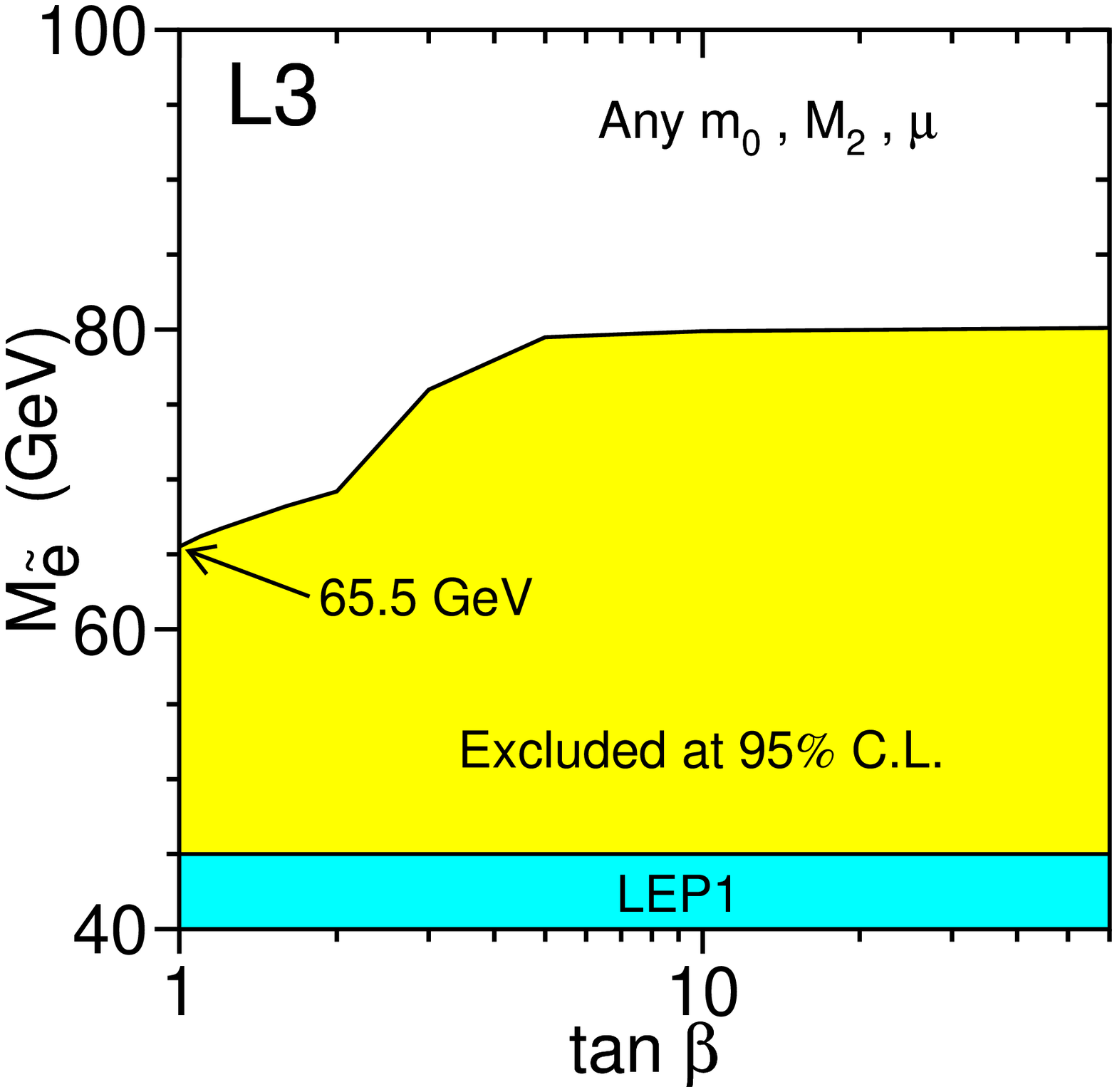}}
  \caption{Lower limit on $M_{\susy{e}_R}$
as a function of $\tan\beta$ and for any value of $m_0$, $M_2$, and
$\mu$. This limit is obtained with 
searches for acoplanar electrons at
centre-of-mass energies between 130 \gev{} and 189 \gev, and single
electrons at 189 \gev.
  \label{fig:mass_sel}} 
\psfig{file=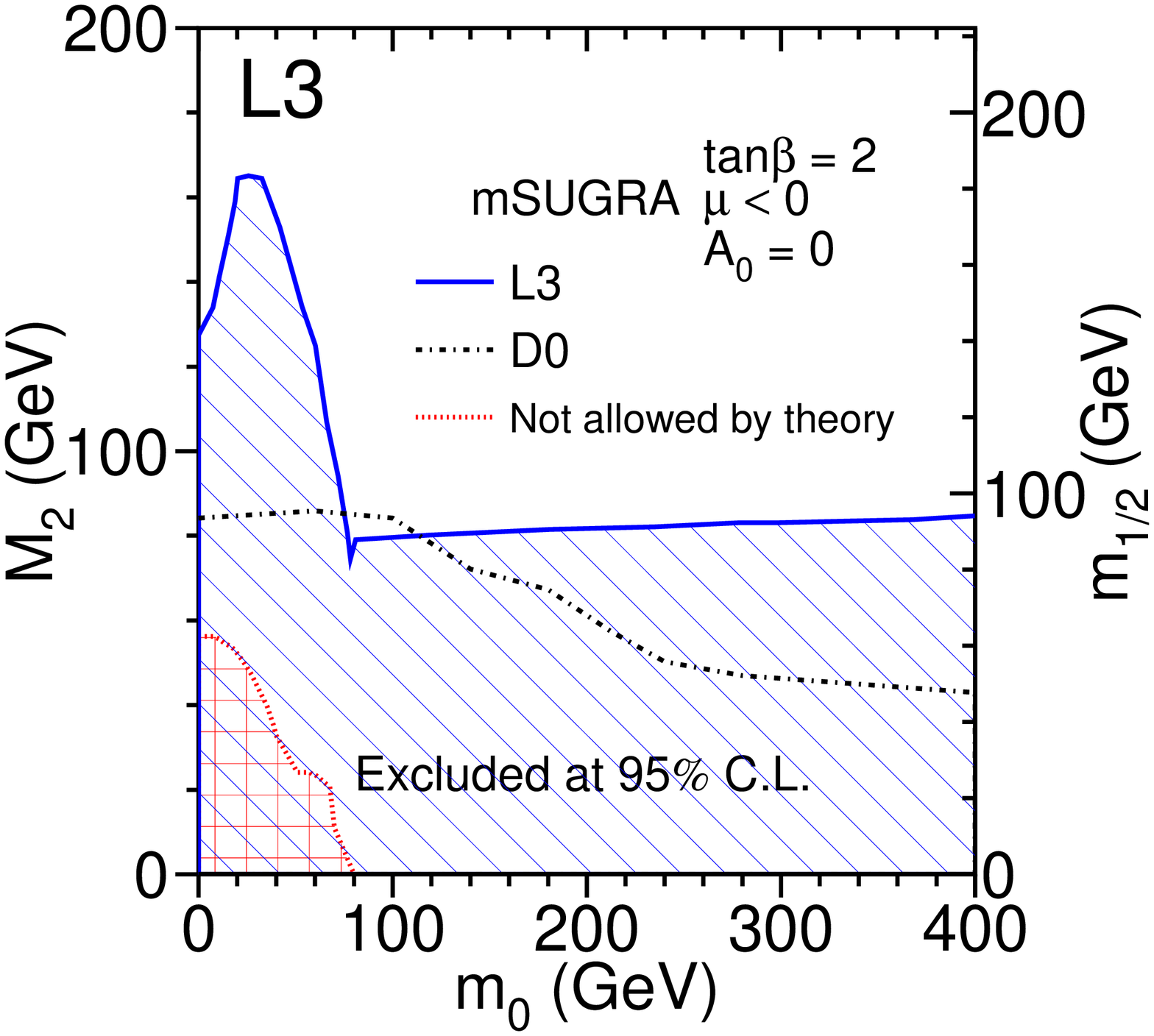,width=10.0 cm} 
 \caption{Exclusion domains in the $M_2-m_0$ plane in the minimal SUGRA framework
for $A_0=0$, $\tan\beta=2$ and $\mu<0$.
The exclusions are obtained by combining scalar electron and muon searches with
chargino and neutralino searches.  The exclusion obtained by D0, at
the Tevatron, from a search for gluinos and scalar 
quarks \protect\cite{d0glui} is also shown. 
The small region in the bottom left corner is theoretically forbidden
within mSUGRA.          \label{fig:m0m2}}
\end{center}

\end{figure}

The limiting factor towards an absolute limit on the scalar electron
mass was the lack of detection efficiency for very small \dm{} values.
This can be overcome in the constrained MSSM by taking profit of the
$\epem\ra\susy{e}_R\susy{e}_L$ process. 
%The lower limit on $M_{\susy{e}_R}$ as a function of $\tan\beta$ 
%and for any value of $m_0$, $M_2$, and
%$\mu$ is shown in Figure~\ref{fig:mass_sel}. This limit is obtained with 
The searches for acoplanar electrons at
centre-of-mass energies between 130 \gev{} and 189 \gev, and single
electrons at $\sqrt{s}=189 \gev$ have been combined to derive a
 lower limit on $M_{\susy{e}_R}$ as a function of $\tan\beta$ 
and for any value of $m_0$, $M_2$ and
$\mu$ as shown in Figure~\ref{fig:mass_sel}.
The new lower limit for the lightest scalar electron independent of
the MSSM parameters, found at $\tan\beta=1$, is:
$$  M_{\susy{e}_R} \geq 65.5 \gev. $$
Assuming a common mass for the scalar leptons at the GUT scale, this limit
holds also for the lightest scalar muon, $\susy{\mu}_R$.

Mass limits on scalar electrons and muons can also be
expressed in terms of the $M_2$ and $m_0$ parameters. This is shown
in Figure \ref{fig:m0m2} where exclusion domains in the $M_2 - m_0$
plane are determined in the minimal supergravity
framework for $A_0=0$, $\tan\beta=2$ and $\mu<0$.   
The exclusion regions
in Figure \ref{fig:m0m2} are obtained by combining scalar electron and
muon searches with chargino and neutralino searches
\cite{charg99ref}. The two
contributions are well separated, as the contribution from scalar
lepton searches is dominant for $m_0 \simle 70 \gev$ while that from 
chargino and neutralino is dominant for $m_0 \simge 70 \gev$.
% In Figure \ref{fig:m0m2} is also shown for comparison the exclusion
% obtained by D0, at the Tevatron, from a search for gluinos and scalar quarks
% \cite{d0glui}. 

%%%%%%%%%%%%%%%%%%%%%%%%%%%%%%%%%%%
%                                 %
%       Acknowledgments           %
%                                 %
%%%%%%%%%%%%%%%%%%%%%%%%%%%%%%%%%%%

\section*{Acknowledgments}

We  express our gratitude to the CERN accelerator divisions for the
excellent performance of the LEP machine. We also acknowledge
and appreciate the effort of the engineers, technicians and support staff 
who have participated in the construction and maintenance of this experiment.

\newpage

%%%%%%%%%%%%%%%%%%%%%%%%%%%%%%%%%%%%%%%%%%%%%%%%%%%%%%%%%%%%%%%%%%%%%%%%%%%%%%%%
%                              BIBLIOGRAPHY
%%%%%%%%%%%%%%%%%%%%%%%%%%%%%%%%%%%%%%%%%%%%%%%%%%%%%%%%%%%%%%%%%%%%%%%%%%%%%%%%
%%\clearpage

\bibliographystyle{l3stylem}

%
%%%%%%%%%%%%%%%%%%%%%%%%%%%%%%%%%%%%%%%%%%%%%%%%%%%%%%%%%%%%%%%%%%%%%%%%%%%%%%
% Author List
%%%%%%%%%%%%%%%%%%%%%%%%%%%%%%%%%%%%%%%%%%%%%%%%%%%%%%%%%%%%%%%%%%%%%%%%%%%%%%
%
\newpage
\typeout{   }     
\typeout{Using author list for paper 186 -?}
\typeout{$Modified: Fri Sep 10 08:43:14 1999 by clare $}
\typeout{!!!!  This should only be used with document option a4p!!!!}
\typeout{   }
%
%
%
%  L A T E X  version!!
%
%
% Make sure that the Lep package has been used!
%\input{Lep.sty}%
%
%\ifx\LepCalled\undefined%
%\typeout{     }%
%\typeout{!!!!!!!!!!!!!!!!!!!!!!!!!!!!!!!!!!!!!!!!!!!!!!!!!!!!!!!!!!!}%
%\typeout{Yikes.  You haven't used the Lep package!}%
%\typeout{Please put \protect\usepackage\protect{Lep\protect} in your preamble,
%         followed by}%
%\typeout{\protect\Lep\protect{1\protect} or \protect\Lep\protect{2\protect}}%
%\typeout{     }%
%\typeout{For now you will get a Lep phase 2 authorlist (may not be right!).}%
%\typeout{!!!!!!!!!!!!!!!!!!!!!!!!!!!!!!!!!!!!!!!!!!!!!!!!!!!!!!!!!!!}%
%\typeout{     }%
%\Lep{2}\fi%

\newcount\tutecount  \tutecount=0
\def\tutenum#1{\global\advance\tutecount by 1 \xdef#1{\the\tutecount}}
\def\tute#1{$^{#1}$}
\tutenum\aachen            % 1
\tutenum\nikhef            % 2
\tutenum\mich              % 3
\tutenum\lapp              % 4
\tutenum\basel             % 5
\tutenum\lsu               % 6
\tutenum\beijing           % 7
\tutenum\berlin            % 8
\tutenum\bologna           % 9 
\tutenum\tata              % 10
\tutenum\ne                % 11
\tutenum\bucharest         % 12
\tutenum\budapest          % 13
\tutenum\mit               % 14 
\tutenum\debrecen          % 15
\tutenum\florence          % 16
\tutenum\cern              % 17 
\tutenum\wl                % 18 
\tutenum\geneva            % 19
\tutenum\hefei             % 20
\tutenum\seft              % 21
\tutenum\lausanne          % 22
\tutenum\lecce             % 23
\tutenum\lyon              % 24
\tutenum\madrid            % 25
\tutenum\milan             % 26
\tutenum\moscow            % 27
\tutenum\naples            % 27
\tutenum\cyprus            % 29
\tutenum\nymegen           % 30
\tutenum\caltech           % 31
\tutenum\perugia           % 32
\tutenum\cmu               % 33
\tutenum\prince            % 34
\tutenum\rome              % 35
\tutenum\peters            % 36
\tutenum\salerno           % 37
\tutenum\ucsd              % 38
\tutenum\santiago          % 39
\tutenum\sofia             % 40
\tutenum\korea             % 41
\tutenum\alabama           % 42
\tutenum\utrecht           % 43
\tutenum\purdue            % 44
\tutenum\psinst            % 45
\tutenum\zeuthen           % 46
\tutenum\eth               % 47
\tutenum\hamburg           % 48
\tutenum\taiwan            % 49
\tutenum\tsinghua          % 50
{
\parskip=0pt
\noindent
{\bf The L3 Collaboration:}
\ifx\selectfont\undefined%  old style font selection
 \baselineskip=10.8pt
 \baselineskip\baselinestretch\baselineskip
 \normalbaselineskip\baselineskip
 \ixpt
\else%                      new style font selection
 \fontsize{9}{10.8pt}\selectfont
\fi
\medskip
\tolerance=10000
\hbadness=5000
\raggedright
\hsize=162truemm\hoffset=0mm
\def\r{\rlap,}
\noindent

M.Acciarri\r\tute\milan\
P.Achard\r\tute\geneva\ 
O.Adriani\r\tute{\florence}\ 
M.Aguilar-Benitez\r\tute\madrid\ 
J.Alcaraz\r\tute\madrid\ 
G.Alemanni\r\tute\lausanne\
J.Allaby\r\tute\cern\
A.Aloisio\r\tute\naples\ 
M.G.Alviggi\r\tute\naples\
G.Ambrosi\r\tute\geneva\
% ...
\newpage
%\rule{\textwidth}{0.4pt}
\begin{list}{A}{\itemsep=0pt plus 0pt minus 0pt\parsep=0pt plus 0pt minus 0pt
                \topsep=0pt plus 0pt minus 0pt}
\item[\aachen]
 I. Physikalisches Institut, RWTH, D-52056 Aachen, FRG$^{\S}$\\
 III. Physikalisches Institut, RWTH, D-52056 Aachen, FRG$^{\S}$
\item[\nikhef] National Institute for High Energy Physics, NIKHEF, 
     and University of Amsterdam, NL-1009 DB Amsterdam, The Netherlands
\item[\mich] University of Michigan, Ann Arbor, MI 48109, USA
\item[\lapp] Laboratoire d'Annecy-le-Vieux de Physique des Particules, 
     LAPP,IN2P3-CNRS, BP 110, F-74941 Annecy-le-Vieux CEDEX, France
\item[\basel] Institute of Physics, University of Basel, CH-4056 Basel,
     Switzerland
\item[\lsu] Louisiana State University, Baton Rouge, LA 70803, USA
\item[\beijing] Institute of High Energy Physics, IHEP, 
  100039 Beijing, China$^{\triangle}$ 
\item[\berlin] Humboldt University, D-10099 Berlin, FRG$^{\S}$
\item[\bologna] University of Bologna and INFN-Sezione di Bologna, 
     I-40126 Bologna, Italy
\item[\tata] Tata Institute of Fundamental Research, Bombay 400 005, India
\item[\ne] Northeastern University, Boston, MA 02115, USA
\item[\bucharest] Institute of Atomic Physics and University of Bucharest,
     R-76900 Bucharest, Romania
\item[\budapest] Central Research Institute for Physics of the 
     Hungarian Academy of Sciences, H-1525 Budapest 114, Hungary$^{\ddag}$
\item[\mit] Massachusetts Institute of Technology, Cambridge, MA 02139, USA
\item[\debrecen] Lajos Kossuth University-ATOMKI, H-4010 Debrecen, Hungary$^\P$
\item[\florence] INFN Sezione di Firenze and University of Florence, 
     I-50125 Florence, Italy
\item[\cern] European Laboratory for Particle Physics, CERN, 
     CH-1211 Geneva 23, Switzerland
\item[\wl] World Laboratory, FBLJA  Project, CH-1211 Geneva 23, Switzerland
\item[\geneva] University of Geneva, CH-1211 Geneva 4, Switzerland
\item[\hefei] Chinese University of Science and Technology, USTC,
      Hefei, Anhui 230 029, China$^{\triangle}$
\item[\seft] SEFT, Research Institute for High Energy Physics, P.O. Box 9,
      SF-00014 Helsinki, Finland
\item[\lausanne] University of Lausanne, CH-1015 Lausanne, Switzerland
\item[\lecce] INFN-Sezione di Lecce and Universit\'a Degli Studi di Lecce,
     I-73100 Lecce, Italy
\item[\lyon] Institut de Physique Nucl\'eaire de Lyon, 
     IN2P3-CNRS,Universit\'e Claude Bernard, 
     F-69622 Villeurbanne, France
\item[\madrid] Centro de Investigaciones Energ{\'e}ticas, 
     Medioambientales y Tecnolog{\'\i}cas, CIEMAT, E-28040 Madrid,
     Spain${\flat}$ 
\item[\milan] INFN-Sezione di Milano, I-20133 Milan, Italy
\item[\moscow] Institute of Theoretical and Experimental Physics, ITEP, 
     Moscow, Russia
\item[\naples] INFN-Sezione di Napoli and University of Naples, 
     I-80125 Naples, Italy
\item[\cyprus] Department of Natural Sciences, University of Cyprus,
     Nicosia, Cyprus
\item[\nymegen] University of Nijmegen and NIKHEF, 
     NL-6525 ED Nijmegen, The Netherlands
\item[\caltech] California Institute of Technology, Pasadena, CA 91125, USA
\item[\perugia] INFN-Sezione di Perugia and Universit\'a Degli 
     Studi di Perugia, I-06100 Perugia, Italy   
\item[\cmu] Carnegie Mellon University, Pittsburgh, PA 15213, USA
\item[\prince] Princeton University, Princeton, NJ 08544, USA
\item[\rome] INFN-Sezione di Roma and University of Rome, ``La Sapienza",
     I-00185 Rome, Italy
\item[\peters] Nuclear Physics Institute, St. Petersburg, Russia
\item[\salerno] University and INFN, Salerno, I-84100 Salerno, Italy
\item[\ucsd] University of California, San Diego, CA 92093, USA
\item[\santiago] Dept. de Fisica de Particulas Elementales, Univ. de Santiago,
     E-15706 Santiago de Compostela, Spain
\item[\sofia] Bulgarian Academy of Sciences, Central Lab.~of 
     Mechatronics and Instrumentation, BU-1113 Sofia, Bulgaria
\item[\korea] Center for High Energy Physics, Adv.~Inst.~of Sciences
     and Technology, 305-701 Taejon,~Republic~of~{Korea}
\item[\alabama] University of Alabama, Tuscaloosa, AL 35486, USA
\item[\utrecht] Utrecht University and NIKHEF, NL-3584 CB Utrecht, 
     The Netherlands
\item[\purdue] Purdue University, West Lafayette, IN 47907, USA
\item[\psinst] Paul Scherrer Institut, PSI, CH-5232 Villigen, Switzerland
\item[\zeuthen] DESY, D-15738 Zeuthen, 
     FRG
\item[\eth] Eidgen\"ossische Technische Hochschule, ETH Z\"urich,
     CH-8093 Z\"urich, Switzerland
\item[\hamburg] University of Hamburg, D-22761 Hamburg, FRG
\item[\taiwan] National Central University, Chung-Li, Taiwan, China
\item[\tsinghua] Department of Physics, National Tsing Hua University,
      Taiwan, China
\item[\S]  Supported by the German Bundesministerium 
        f\"ur Bildung, Wissenschaft, Forschung und Technologie
\item[\ddag] Supported by the Hungarian OTKA fund under contract
numbers T019181, F023259 and T024011.
\item[\P] Also supported by the Hungarian OTKA fund under contract
  numbers T22238 and T026178.
\item[$\flat$] Supported also by the Comisi\'on Interministerial de Ciencia y 
        Tecnolog{\'\i}a.
\item[$\sharp$] Also supported by CONICET and Universidad Nacional de La Plata,
        CC 67, 1900 La Plata, Argentina.
\item[$\diamondsuit$] Also supported by Panjab University, Chandigarh-160014, 
        India.
\item[$\triangle$] Supported by the National Natural Science
  Foundation of China.
\item[\dag] Deceased.
\end{list}
}
\vfill

%%% Local Variables: 
%%% mode: latex
%%% TeX-master: t
%%% End:

%%% Local Variables: 
%%% mode: latex
%%% TeX-master: t
%%% TeX-master: t
%%% TeX-master: t
%%% End: 

%%% Local Variables: 
%%% mode: latex
%%% TeX-master: t
%%% End: 

%%% Local Variables: 
%%% mode: latex
%%% TeX-master: t
%%% End: 

%%% Local Variables: 
%%% mode: latex
%%% TeX-master: t
%%% End: 

%\newpage
\clearpage
%

%%%%%%%%%%%%%%%%%%%%%%%%%%%%%%%%%%%%%%%%%%%%%%%%%%%%%%%%%%%%%%%%%%%%%%%%%%%%%%%%
%                              Tables
%%%%%%%%%%%%%%%%%%%%%%%%%%%%%%%%%%%%%%%%%%%%%%%%%%%%%%%%%%%%%%%%%%%%%%%%%%%%%%%%
\clearpage

\begin{table} [htbp!]
\begin{center} 
\begin{tabular}{|l|c|c|c|c|c|} \hline         
\multicolumn{6}{|c|}{Scalar electron selections } \\  \hline
  \multicolumn{2}{|c|}{\dm\, (\gev)}  & \multicolumn{2}{|c|}{$3-5$}  & $10-40$ & $60-90$  \\  \hline
 $E_{lep}$ (\gev)    & $\le$ & \multicolumn{2}{|c|}{5.34}  & 37.4   & 59.8  \\  \hline
 $\sum E_{lep}$ (\gev) & $\ge$ & \multicolumn{2}{|c|}{4.45}  & 16.9 & 65.6  \\  \hline
 $E_{vis}/\sqrt{s}$  & $\le$  & \multicolumn{2}{|c|}{0.12}  & 0.36   & 0.63  \\  \hline
 $p_{\perp}$ (\gev)  & $\ge$  & \multicolumn{2}{|c|}{3.62} & 1.45   & 8.95   \\  \hline
  Acollinearity (rad) & $\le$ & \multicolumn{2}{|c|}{2.26} & 3.10   &  --  \\  \hline
  Acoplanarity  (rad) & $\le$ & \multicolumn{2}{|c|}{2.95} & 3.08   &  3.01 \\  \hline
 $E^{\perp}_{25}$ (\gev)& $\le$ & \multicolumn{2}{|c|}{--} & 3.8  & 7.51   \\  \hline
 $E_{25}$ (\gev)   &$\le$   &  \multicolumn{2}{|c|}{0.28} & 3.2   & 3.52   \\  \hline
 $E_{60}^b$ (\gev)  &$\le$  &  \multicolumn{2}{|c|}{2.93} & 3.7   & 4.59   \\  \hline
 sin($\theta_{miss}$)& $\ge$ & \multicolumn{2}{|c|}{0.46} & 0.60  &  0.20  \\  \hline
 $E_{TTL}$ (\gev)   & $\ge$  &\multicolumn{2}{|c|}{ 3.22} & 3.97  &  2.70   \\  \hline
\multicolumn{6}{|c|}{Scalar muon selections} \\  \hline
 \multicolumn{2}{|c|}{\dm\, (\gev)} &  \multicolumn{2}{|c|}{$3-5$} & $10-40$ & $60-90$ \\  \hline
 $E_{lep}$ (\gev)  & $\le$  &\multicolumn{2}{|c|}{9.97}  & 25.6  & 78.4   \\  \hline
 $E_{vis}/\sqrt{s}$ & $\le$ & \multicolumn{2}{|c|}{0.16} & 0.30  & 0.58   \\  \hline
 $p_{\perp}$ (\gev)  & $\ge$ & \multicolumn{2}{|c|}{2.69} & 8.53 &  11.2  \\  \hline
 Acollinearity   (rad) & $\le$ & \multicolumn{2}{|c|}{3.09} & 3.09 &  2.41 \\  \hline
 Acoplanarity    (rad)  & $\le$& \multicolumn{2}{|c|}{2.90} & 3.11 &  2.44 \\  \hline
 $E^{\perp}_{25}$ (\gev)& $\le$  & \multicolumn{2}{|c|}{--} & 3.97 & 4.04  \\  \hline
 $E_{25}$ (\GeV)    &$\le$   &  \multicolumn{2}{|c|}{1.0}  & 2.93   & 3.43  \\  \hline
 $E_{60}^b$ (\GeV)  &$\le$   &  \multicolumn{2}{|c|}{9.94} & 7.79   & 6.67  \\  \hline
 sin($\theta_{miss}$) & $\ge$ &  \multicolumn{2}{|c|}{0.80} & 0.53 & 0.35 \\  \hline
 $E_{TTL}$ (\GeV)      & $\ge$ & \multicolumn{2}{|c|}{2.44}  & 2.35 &  4.99 \\  \hline
\multicolumn{6}{|c|}{Scalar tau selections } \\  \hline
\multicolumn{2}{|c|}{\dm\, (\gev)} & $3-7$ &$7-15$ & $15-30$& $30-90$ \\  \hline
 $E_{vis}/\sqrt{s}$     & $\ge$ & 0.04 & 0.06  & 0.08  & 0.11 \\ \hline
 $E_{vis}$ (\gev)       & $\le$ & 21.9 & 38.1  & 54.4 & 76.1\\ \hline
 $p_{\perp}$ (\gev)      & $\ge$ & 3.68 & 9.43  & 9.12  & 13.7\\ \hline
 $p_{\perp}/E_{vis}$     & $\ge$ & 0.08  & 0.36 & 0.19 & 0.30\\ \hline
 Acollinearity (rad)    & $\le$ & 3.08   & 2.98  & 3.14 &3.03\\ \hline
 Acoplanarity  (rad)    & $\le$ & 3.13   & 3.08  & 3.07  &2.77\\ \hline
 sin($\theta_{miss}$)     & $\ge$ & 0.85  & 0.67 & 0.58 & 0.55\\ \hline
 $E_{25}^{\perp}$ (\gev) & $\le$ & 8.97   & 7.24  & 1.56 & 0.87\\ \hline
 $E_{TTJ}$ (\gev)       & $\ge$ & 2.14  & 2.23 & 3.81 & 0.89\\ \hline
 $E_{TTJ}/p_{\perp}$     & $\ge$ & 0.21  & 0.13 & 0.13 & 0.04\\ \hline
 Max track acoplanarity  (rad)  & $\le$ & 2.98   & 2.97  & 2.93   &2.66\\ \hline
 $y_{\perp}$             & $\le$ & 0.38  & 0.33 & 0.36 & 0.73\\ \hline
 $E^{\ell}$ (\gev)      & $\le$ & 14.3   & 33.8  & 50.8  & 62.2\\  \hline

%\multicolumn{5}{|c|}{Scalar tau selections } \\  \hline
% $E_{lep}$  (GeV)       & $\le$ & 16.1    & 29.9       & 36.2      \\  \hline
%% $E_{lep}$  (GeV)       & $\ge$ & 2.32    & 14.0       & 11.8     \\  \hline
% $E_{vis}/\sqrt{s}$      & $\le$ & 0.10    & 0.24       & 0.28    \\  \hline
% $p_{\perp}$ (GeV)      & $\ge$ &  7.92    & 10.3      &  9.55   \\  \hline
%  Acollinearity (rad)    & $\le$ &  3.10   & 2.66      &  2.63    \\  \hline
%  Acoplanarity (rad)     & $\le$ &  2.76   & 2.71      &  2.34    \\  \hline
%  $E^{\perp}_{25}$ (GeV) & $\le$ &  --     & 2.08      & 1.13     \\  \hline
%  $\TRE$ (rad)           &$\le$  &   --     & 5.83     & --     \\  \hline
%  $E_{60}^b$ (GeV)       &$\le$  &  --     & 3.61      & 1.43     \\  \hline
%  sin($\theta_{miss}$)   & $\ge$ &  0.77   & 0.66      & 0.53     \\  \hline
%  $E_{TTL}$ (GeV)        & $\ge$ &  1.21    & 2.01     & 3.51    \\  \hline
\end{tabular} 
\end{center} 
\caption{Optimised cut values for the scalar lepton
         searches for the different \dm\, ranges. They
        are determined with the
         optimisation procedure described 
in Section~\protect\ref{sec:optimization}.}
\label{tab1}
\end{table}   

\begin{table} [htbp!]
\begin{center} 
\begin{tabular}{|c|c|c|c|c|c|c|c|c|} \hline  
 & \multicolumn{2}{|c|}{Low \dm}& \multicolumn{2}{|c|}{Medium \dm}& 
\multicolumn{2}{|c|}{High \dm}& \multicolumn{2}{|c|}{Combined} \\  \hline
  & $N_{data}$ & $N_{exp}$ & $N_{data}$ & $N_{exp}$ & $N_{data}$ & $N_{exp}$ 
& $N_{data}$ & $N_{exp}$ \\  \hline
\susy{e}   & 7  & 6.0 & 3 & 4.8 & 11 & 12.4 & 21  & 23.0 \\ \hline
\susy{\mu} & 10  & 11.5 & 2 & 1.0 & 8 & 9.1 & 19  & 21.0 \\ \hline
\susy{\tau} & 23  & 23.1 & 5 & 7.5 & 33 & 29.4 & 56 & 51.6 \\ \hline
%Combined & redo & redo & redo & redo & redo & redo & redo & redo \\ \hline
\end{tabular} 
\caption[cascade]{Results of the acoplanar lepton searches:
           $N_{data}$ is the number of observed events and
           $N_{exp}$ is the number of expected events from Standard 
           Model processes for the total integrated
           luminosity collected at $\sqrt{s}=189 \gev$. 
\label{tab8}}
\end{center} 
\end{table} 

%%% \CHECK Should we quote one digit less?
\begin{table} [htbp]
\begin{center} 
\begin{tabular}{|c|c|c|c|c|c|c|} \hline
     & \multicolumn{6}{|c|}{$\sqrt{s} =189 \gev$}   \\ \cline{2-7}
     & \multicolumn{2}{c|}{$M_{\selectron{\pm}} = 90 \gev$}
     & \multicolumn{2}{c|}{$M_{\smuon{\pm}}     = 80 \gev$}
     & \multicolumn{2}{c|}{$M_{\stau{\pm}}      = 70 \gev$} 
            \\ \hline
$\ee \rightarrow$  &\multicolumn{2}{|c|}{~~\selectron{\pm}\selectron{\mp}~~} 
     &\multicolumn{2}{|c|}{~~\smuon{\pm}\smuon{\mp}~~ }
     &\multicolumn{2}{|c|}{~~\stau{\pm}\stau{\mp}~~ } 
            \\  \hline
{~~ \dm (\GeV)~~}  & ~~$\epsilon $ (\%) & $N_{exp}$ &
                    ~~$\epsilon $ (\%) & $N_{exp}$ &
                    ~~$\epsilon $ (\%) & $N_{exp}$   \\
\hline
3   & 20.4   & 2.3   & 11.5   & 11.5  & 1.4   & 23.1     \\
5   & 18.7   & 5.9  & 24.0   & 12.3  & 6.4    & 23.1   \\
10  & 44.5   & 4.8    & 33.3   & 1.0 & 9.1  & 7.5     \\
20  & 53.8   & 4.8    & 32.1   & 1.0  & 26.1  & 16.5     \\
30  & 49.1   & 4.8   & 35.6   & 9.7  & 26.3  & 16.5     \\
40  & 54.4   & 16.6   & 33.4   & 9.1  & 30.0 & 29.4     \\
50  & 57.9   & 16.6   & 33.1   & 9.1  & 28.2  & 29.4     \\
60  & 56.1   & 11.9    & 31.6   & 9.1  & 29.1  & 29.4       \\
68  & 55.9  & 11.9    & 29.9    & 9.1   & 29.7   & 24.4       \\
78  & 55.9   & 11.9    & 27.2     & 9.1    & --    & --       \\
88  & 53.4   & 11.9    & --     & --    & --    & --       \\

\hline 
\end{tabular} 
\caption[cascade]{Scalar electron, muon and tau efficiencies ($\epsilon$) and
           number of events expected from Standard Model processes
           ($N_{exp}$).  
           Results at $\sqrt{s}=189 \gev$ are listed as a function of
           $\dm$ for different
           $M_{\slepton{\pm}}$ values.}
\label{tab5}
\end{center} 
\end{table}

\pagebreak 

\end{document}